**Title:**

Efficient Beam Manipulation with Phase Symmetry Operations on Modulated Metasurfaces

**Authors:**

*Yang Cai*, *Peng Mei\**, *Gert Frølund Pedersen*, *Shuai Zhang*

Antennas, Propagation and Millimeter-wave System (APMS) section, Department of Electronic System, Aalborg University, Aalborg, 9220, Denmark.

Corresponding authors: *Peng Mei*, mei@es.aau.dk

**Abstract**

Beam manipulation is of paramount importance in wave engineering, enabling diverse beam shapes like pencil beams, flat-top beams, and isoflux beams to cater to various application missions. Among the beams, shaping flat-top and isoflux beams remains challenging with the traditional synthesis approaches that mainly rely on optimization algorithms. Here, we develop modulated metasurfaces to efficiently generate flat-top and isoflux beams from the first principle of field superposition with negligible optimizations, by performing phase symmetry operations. The theoretical analysis not only facilitates the shaping of 1D and 2D flat-top and isoflux beams but also exhibits controllable beamwidths. Experimental validation confirms the efficacy of the phase symmetry operations in generating flat-top beams with adjustable beamwidths. The concept of phase symmetry operations can be extended to other vectorial components, offering potential applications for the manipulation of various wave types such as acoustic waves, water surface waves, and beyond, thereby advancing related applications.

**Introduction**

The rapid advancement of wireless technologies, increasingly intricate electromagnetic environments, and various application missions impose more stringent requirements on antennas which are vital components to receive and transmit electromagnetic waves. Specifically, in the realm of long-distance and point-to-point wireless communications, high-gain and pencil-beam antennas are favored for their ability to compensate for the intrinsic free space path loss, thereby permitting a robust link budget. In contrast, the flat-top beam which provides homogeneous and uniform power density within the main beam and exhibits a sharp cutoff and low side lobes simultaneously is demanded in various applications including precisely sensing[1], lithography[2,3], optical wireless communication (OWC)[4]. Moreover, in the context of space-to-ground satellite communications, isoflux beams have proven to be highly advantageous considering the spherical shape of the Earth as they offer maximum signal coverage and maintain a robust link budget[5,6]. Conventional antenna arrays excel in providing high gain and pencil-beam radiation characteristics through synchronization of all the radiating elements in phase. If stringent specifications on sidelobe levels are considered, amplitude modulation should be additionally introduced to the conventional antenna arrays such as Chebyshev and Taylor synthesis techniques, which have been extensively studied in the past. In contrast, the synthesis of flat-top and isoflux radiation beams using conventional antenna arrays remains relatively understudied and underreported. The classic method to synthesize flat-top beams is based on the Woodward-Lawson sampling technique, which can quantitively specify the amplitude and phase of each antenna element of a linear antenna array[7,8]. Nevertheless, the Woodward-Lawson sampling synthesis method is not suitable for practical implementation of large-scale linear arrays since the calculated magnitude of each antenna element is challenging to realize and sometimes unfeasible. Another prevalent synthesis method in the literature relies heavily on optimization algorithms, where the amplitude and phase distribution across the array are optimized and determined based on predefined goals and objectives [9–14]. While global optimization methods have proven to be effective for many synthesis problems, they often entail complex, time-consuming, and resource-intensive iterative processes, particularly when applied to the synthesis of large-scale arrays. Additionally, practical implementation of such large-scale arrays poses significant challenges, since designing power dividers to achieve the required amplitude modulation can be exceedingly intricate and sometimes unfeasible, as in the case of the Woodward-Lawson sampling technique.

The emergence of metamaterials and metasurfaces has introduced new paradigms for manipulating a wide range of waves such as acoustics waves[15–17], electromagnetic (EM) waves[18–23], and visible light[24–26] by locally controlling the responses of the unit cells comprising the metamaterials and metasurfaces. Metasurfaces, being two-dimensional (2D) metamaterials, are typically artificially engineered surfaces consisting of periodic or quasi-periodic meta-atoms. They exhibit distinct features, such as low insertion loss, ease of fabrication, and more. Metasurfaces, therefore, find applications in a multitude of application scenarios, including beam manipulation[27–29], polarization conversion[30–33], imaging [34–38] as well as holograms[34–38], and so forth. Additionally, when metamaterials and metasurfaces are spatially illuminated by an external source emitting spherical waves, the electric-field distributions across their surfaces exhibit intrinsic amplitude modulation, which can be further harnessed to control wave properties.

As previously mentioned, beam manipulation always necessitates the regulation of both amplitude and phase. The intrinsic and flexible amplitude modulation of metamaterials and metasurfaces, which can be modified by simply altering the distance between the external source and the metamaterials or metasurfaces, streamlines wave manipulations to a certain extent compared to the conventional antenna arrays, especially in the practical implementation of large-scale arrays. In this article, we introduce an efficient synthesis method based on phase symmetry operations to generate desired beam shapes with modulated metasurfaces [39]. These beam shapes can be flexibly controlled by adjusting just a few key parameters to achieve pencil, flat-top, or isoflux beams that have a wide range of applications in point-to-point wireless communications, high density of users, and space-to-ground satellite communications, as illustrated in Fig. 1. While the pencil beam has received extensive attention and study in the literature, methodologies for efficiently synthesizing flat-top and isoflux beams are relatively scarce and predominately rely on optimization algorithms. As a result, we mainly delve into the generation and manipulation of flat-top and isoflux beams with phase symmetry operations on modulated metasurfaces. We first theoretically derive that the fundamental element to generate flat-top and isoflux beams is a quasi-sawtooth-shaped beam based on the first principle of field superposition. The quasi-sawtooth-shaped beam can be generated by a subordinate semi-metasurface of a complete metasurface that is illuminated by an offset virtual focus that is characterized with a offset angle of $\theta_{vf}$, and can be flexibly shifted in angular domain with its shape constant by imposing an additional phase gradient $\theta_p$. By performing phase symmetry operations, which involve mirroring the subordinate semi-metasurface to create a complete metasurface, two flipped quasi-sawtooth-shaped beams originating from the two semi-metasurfaces are generated accordingly, leading to a superposed beam eventually. The properties of the superposed beam, such as flat-top or isoflux characteristics, are flexibly regulated and controlled by the angle pair ($\theta_{vf}$, $\theta_p$). Similarly, 2D flat-top or isoflux beams can also be achieved by performing rotational phase symmetry operations, which rotate the subordinate sequence concerning the center of the metasurface. Furthermore, we analyze the upper limit of the achievable beamwidth of flat-top beams for a well-defined metasurface. We provide theoretical and experimental evidence demonstrating that phase symmetry operations can efficiently regulate the beams of modulated metasurfaces. We anticipate that the proposed phase symmetry operations will enrich the manipulation techniques in wave engineering, advancing a broad range of applications in this field.

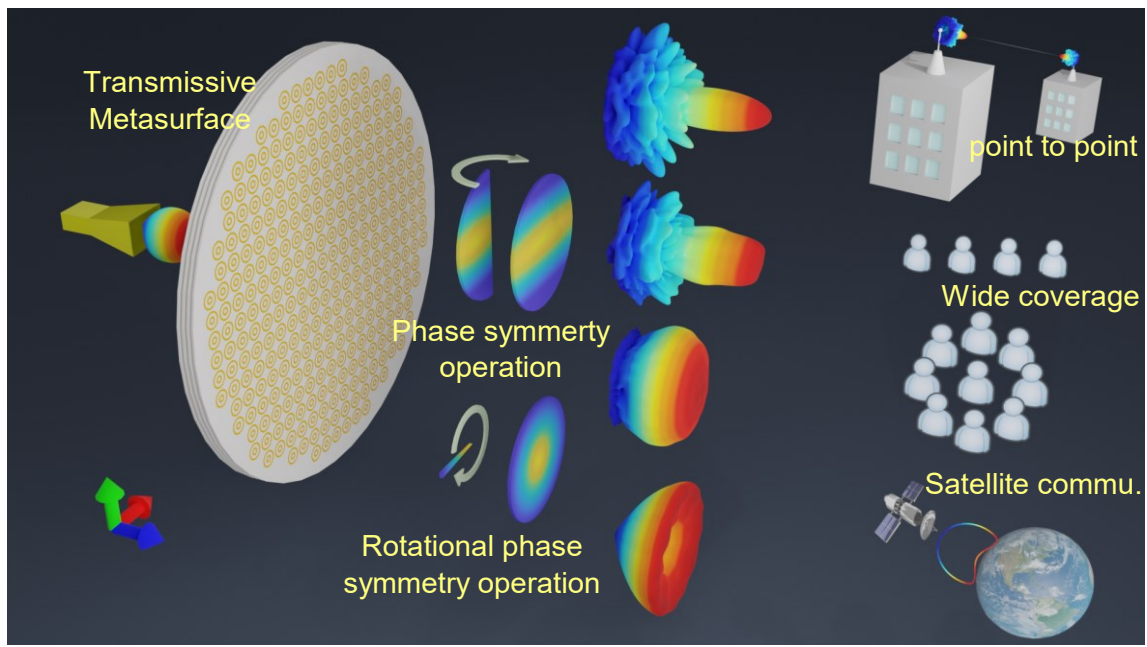

**Fig. 1 |** The various radiation beams generated by modulated metasurfaces based on phase symmetry operations and rotational phase symmetry operations. These beams have their respective applications such as long-distance point-to-point wireless communications, wide coverage wireless communications with one-/two-dimensional high density of users, and space-to-ground satellite communications.

## Results

### Generation of Quasi-Sawtooth-Shaped Beam

The fundamental principle that underlies the creation of flat-top and isoflux beams relies on vectorial superpositions of electric fields, which is originally derived from the superposition of two sawtooth functions as explained in Supplementary Note 1. One prevalent solution to generate sawtooth-shaped beams is based on optimization algorithms that optimize the amplitude and phase of each radiating element to fit the defined objectives. Here, we elaborate on the generation of a sawtooth-shaped beam using a modulated metasurface that is spatially illuminated by an external source, without the need for optimization algorithms.

To quantitatively describe the radiation characteristics of a metasurface, a mathematical expression of the radiation pattern is formulated. First of all, the radiation pattern of an external source is typically characterized with $\cos^{q_f}\theta$ ($q_f$ = 2 is assumed in the following calculation and the q-factor of the meta-atom $q_e$ = 1 is assumed similarly). Therefore, the radiation pattern originating from a metasurface can be formulated as [15–17]

$$E(\theta,\varphi) = \sum_{m=1}^{M}\sum_{n=1}^{N} \cos^{q_e}\theta \frac{\cos^{q_f}\theta_f(m,n)}{|\vec{r}_{mn}-\vec{r}_f|} \cdot e^{-jk(|\vec{r}_{mn}-\vec{r}_f|-\vec{r}_{mn}\cdot\hat{u})} \cdot |T_{mn}|e^{j\psi_{mn}}, \tag{1}$$

where $\vec{r}_{mn}$ and $\vec{r}_f$ is the position vector of the $mn^{th}$ meta-atom and the on-axis external source, respectively, as shown in Fig. 2a; k is the wave number in free space; $\hat{u}$ is the unit vector of the observation direction; $|T_{mn}|$ and $\Psi_{mn}$ are the transmission amplitude and transmission phase of the $mn^{th}$ meta-atom, respectively. The term $\cos^{q_f}\theta_f(m,n)/|\vec{r}_{mn}-\vec{r}_f|$ in Eq. (1) exhibits the intrinsic amplitude modulation across the aperture of the metasurface.

When a metasurface is illuminated by an on-axis external source, it can generate a pencil beam at the boresight direction if the phase of the metasurface is adequately compensated. As the external source rotates along an arc with the phase compensation held constant, a tilted pencil beam can then be generated, as depicted in Fig. 2b. The tilted angle of the beam, denoted as $\alpha_1$, is typically smaller than the geometrical angle $\theta_1$, which represents the angle of the external source relative to the center of the metasurface[40].

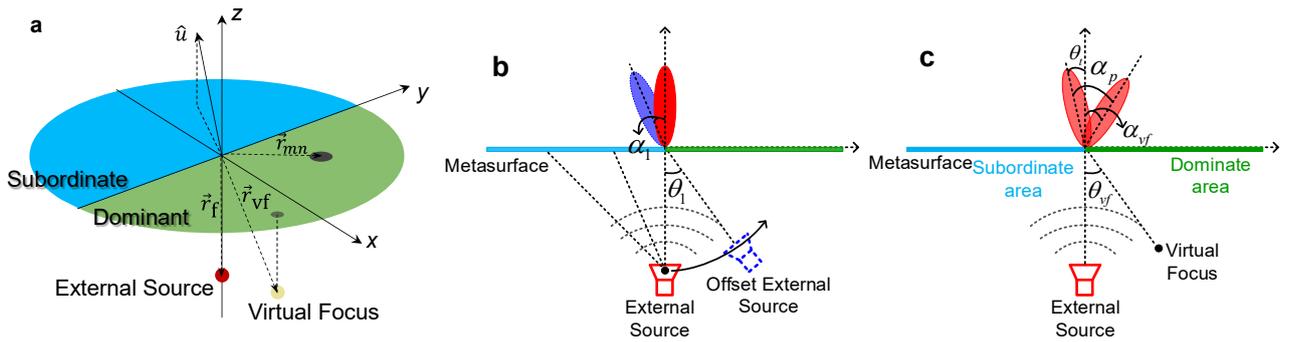

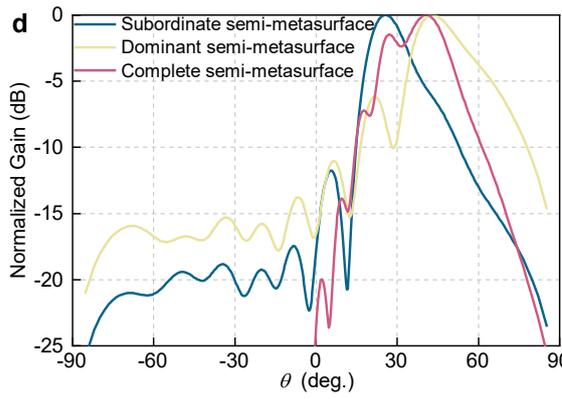

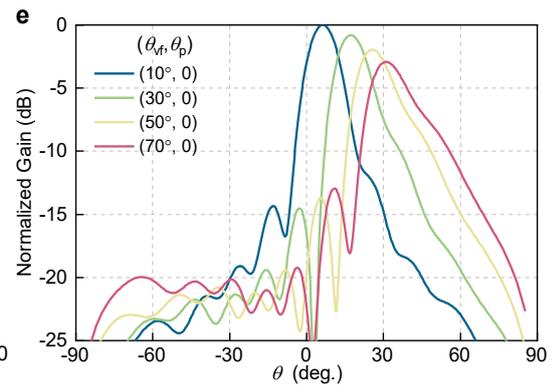

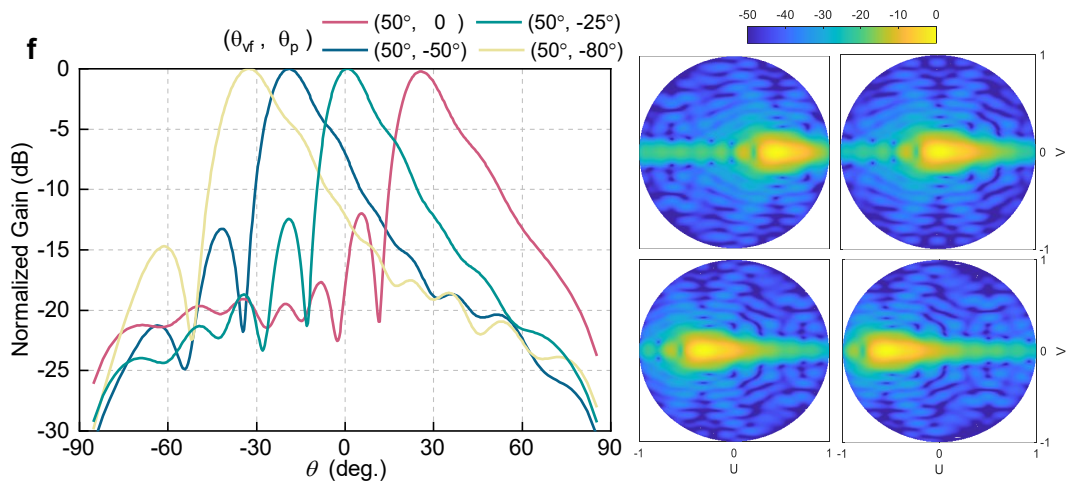

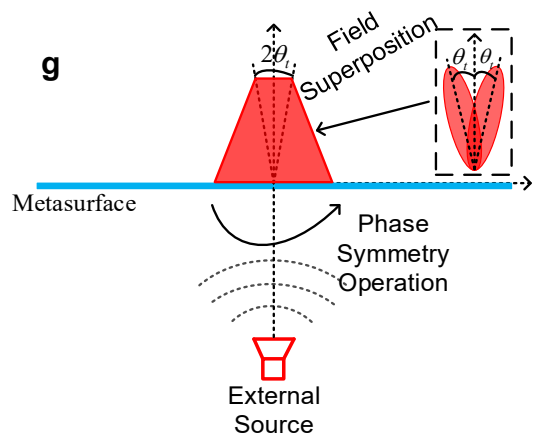

**Fig. 2 | Generation and control of quasi-sawtooth-shaped beam and phase symmetry operations. a.** The diagram of a general metasurface illuminated by an external source. **b.** A metasurface designed to compensate for the spatial phase delay concerning an on-axis external source can steer the beam when the external source is offset. **c.** A metasurface designed to compensate for the spatial phase delay concerning an off-axis external source can steer the beam when the external source is moved to the on-axis. **d.** The radiation patterns originating from subordinate semi-metasurface, dominate semi-metasurface, and complete metasurface. **e.** The radiation pattern originating from the subordinate semi-metasurface with different tilted angles of the virtual focus $\theta_{vf}$. **f.** The quasi-sawtooth-shaped beam originating from the subordinate semi-metasurface with different additional phase gradients $\theta_p$ and $\theta_{vf}$ = 50°. **h.** The diagram of phase symmetry operations to generate flat-top or isoflux beam.

We subsequently displace the external source off-axis with an angle of $\theta_{vf}$ along a defined arc, referred to as a virtual focus in Fig. 2c. By compensating for the spatial phase delay concerning the off-axis virtual focus, the metasurface will yield a pencil beam at the boresight direction when the external source is located at the virtual focus. In contrast, when the same metasurface is illuminated by an on-axis external source along the same arc, equivalent to the reverse process depicted in Fig. 2b, it generates a tilted beam at an angle of $\alpha_{vf}$ as envisioned in Fig. 2c. Geometrically speaking, the metasurface can be spatially divided into two semi surfaces with respect to the virtual focus. For ease of description, the two semi-surfaces are categorized into dominant and subordinate areas based on the levels of illumination power originating from the virtual focus on the surface. The dominant semi-surface constitutes the half-area with higher power levels, into which the projection of the virtual focus falls. In contrast, the subordinate area comprises the other half semi-surface characterized by lower power levels or its position away from the virtual focus as illustrated in Fig. 2c. Then, we investigate the radiation characteristics of the two semi-metasurfaces as well as the complete metasurface for a specific configuration when the external source is positioned on-axis. The specific configuration is: the metasurface consists of 316 meta-atoms forming a circular aperture with a diameter of 20 meta-atoms. The focal distance to aperture diameter ratios (F/D) in this scenario is 0.4 at 28 GHz, the periodicity of the meta-atom is 5 mm, and $\theta_{vf}$ is 50°. Consequently, the radiation pattern resulting from the complete metasurface can be easily calculated using Eq. (1), with the assumption that $|T_{mn}|$ is uniformly set to 1 for all meta-atoms. Fig. 2d plots the radiation pattern of the complete metasurface on the xoz-plane. To determine the radiation pattern associated with each semi-surface, specific constraints are imposed to accurately obtain the radiation characteristics resulting from each semi-surface. With well-defined dominant and subordinate areas, the radiation characteristics of each semi-surface are computed by explicitly setting the amplitude of the other semi-surface to zero. Referring to the diagram in Fig. 2c, this can be expressed mathematically as:

$$\begin{cases} \text{Dominant area:} & |T_{mn}| = 0, \text{ for } x < 0, \\ \text{Subordinate area:} & |T_{mn}| = 0, \text{ for } x > 0, \end{cases} \quad (2)$$

where $|T_{mn}|$ is the transmission amplitude of the $mn^{th}$ meta-atom of the metasurface. Substituting Eq. (2) into Eq. (1), the radiation patterns resulting from each semi-metasurface on the xoz-plane are computed and depicted in Fig. 2d as well. It is evident that the subordinate semi-metasurface can produce a robust quasi-sawtooth-shaped beam. While the dominant semi-metasurface also generates a quasi-sawtooth-shaped beam, it exhibits relatively higher sidelobe levels. The reasons for the distinct radiation beams generated by the dominant and subordinate semi-metasurfaces are comprehensively addressed in Supplementary Note 2. We also investigate the radiation beam resulting from the subordinate semi-metasurface with different tilted angles of the virtual focus $\theta_{vf}$. As seen in Fig. 2e, the titled angle of the quasi-sawtooth-shaped beam consistently increases with the tilted angle of $\theta_{vf}$, and the 6-dB beamwidth of the quasi-sawtooth-shaped beam also widens as $\theta_{vf}$ increases. Moreover, the quasi-sawtooth-shaped beam always remains constant with different $\theta_{vf}$, which is a good metric to facilitate a robust flat-top or isoflux beam since even a slight distortion in the quasi-sawtooth-shaped beam may result in significant ripples in flat-top or isoflux beams.

As mathematically discussed in Supplementary Note 1, to superpose a flat-top or isoflux beam, the quasi-sawtooth-shaped beam should be capable of shifting in the angular domain so as to flexibly control the magnitude of the intersection point at the boresight direction relative to the peak magnitude as shown in Supplementary Fig. S1a. As a result, an additional phase gradient denoted as ($\theta_p$, $\varphi_p$) should be imposed on all meta-atoms, the phase shift ($\Psi_{mn}$) of each meta-atom is thereby modified to:

$$\psi_{mn} = k|\vec{r}_{mn} - \vec{r}_{vf}| - k\vec{r}_{mn} \cdot \hat{u}_p, \tag{3}$$

$$\vec{r}_{vf} = \hat{x}F\sin\theta_{vf}\cos\varphi_{vf} + \hat{y}F\sin\theta_{vf}\sin\varphi_{vf} + \hat{z}F\cos\theta_{vf}, \tag{4}$$

$$\hat{u}_p = \hat{x}\sin\theta_p\cos\varphi_p + \hat{y}\sin\theta_p\sin\varphi_p + \hat{z}\cos\theta_p, \tag{5}$$

where the first term is the spatial phase delay from the virtual focus to the $mn^{th}$ meta-atom, and the second term implies the phase contribution from the additional phase gradient. $F$ represents the spatial distance between the virtual focus and the center of the metasurface. The vectors $\vec{r}_{vf}$ and $\hat{u}_p$ correspond to the position vector of the virtual focus and the unit vector of the direction of the scanned beam, respectively. For the sake of conciseness, the virtual focus is positioned within the *xoz*-plane and the additional progress phase gradient also occurs within the *xoz*-plane in the theoretical analysis and calculations, i.e., $\varphi_{vf} = \varphi_p = 0$. We study a specific case here where $\theta_{vf} = 50°$ and $\theta_p = 0°$, the quasi-sawtooth-shaped beam resulting from the subordinate semi-surface is plotted in Fig. 2f as indicated by the red line. Then, we investigate the effects of the additional phase gradient on the quasi-sawtooth-shaped beam. The shifted quasi-sawtooth-shaped beam on the xoz-plane is calculated by substituting Eqs. (3)-(5) and Eq. (2) to Eq. (1) and depicted in Fig. 2f with $\theta_p$ = -25°, -50°, -80°, respectively. For more features, the corresponding 2D electric-field distributions are also supplied as illustrated in Fig. 2f. It is evident that the quasi-sawtooth-shaped beam can be flexibly shifted to the corresponding direction with different additional phase gradient $\theta_p$ while always keeping its shape well. Hence, for the quasi-sawtooth-shaped beams with varying tilted angles of $\theta_{vf}$ shown in Fig. 2e, the additional phase gradient $\theta_p$ can be selected accordingly to shift the quasi-sawtooth-shaped beam. This enables flexible control of the magnitude of the quasi-sawtooth-shaped beam at the boresight direction (i.e., $\theta = 0°$) for different tilted angles of $\theta_{vf}$. As a result, it is concluded that the tilted angle of the virtual focus $\theta_{vf}$ and the additional phase gradient $\theta_p$ are two critical parameters directly impacting the characteristics of the quasi-sawtooth-shaped beam, ultimately dictating the performance of the superposed flat-top or isoflux beams.

## Generation of a One-dimensional Flat-top Beam

We have generated the desired quasi-sawtooth-shaped beam so far. Additionally, it can be flexibly shifted to control the magnitude of the intersection point at the boresight direction (i.e., $\theta = 0$), which is essential to generate a pencil-, flat-top, or isoflux beam. As mathematically discussed in Supplementary Note 1, a flipped quasi-sawtooth-shaped beam is necessary to superpose with the original quasi-sawtooth-shaped beam to form a flat-top or isoflux beam as illustrated in Supplementary Fig. S1b, c. To facilitate the flipped quasi-sawtooth-shaped beam, phase symmetry operations are executed by mirroring the subordinate semi-metasurface to build a complete metasurface as illustrated in Fig. 2g. The total radiation beam resulting from the complete metasurface is obtained by superposing the radiation beams produced by the two semi-metasurfaces. As a result, the manipulation of the radiation beam resulting from the complete metasurface can be efficiently translated into the regulation of the radiation beam produced by the semi-metasurface. Here, we demonstrate the generation of a flat-top beam, while the production of an isoflux beam is similarly discussed in Supplementary Note 3. We select the angle pair ($\theta_{vf}$, $\theta_p$) of (50º, -50º) since the magnitude of the intersection point at the boresight direction is approximately half of the peak amplitude (equivalent to a normalized gain 6dB lower than the maximal value) as illustrated in Fig. 2f, which is essential to permit a robust flat-top beam as discussed in Supplementary Note 1. We therefore calculate the radiation beams resulting from the two semi-metasurfaces and the complete metasurface, as depicted in Fig. 3a. It is evident that the total radiation beam from the complete metasurface is essentially a quasi-superposition of the two quasi-sawtooth-shaped beams originating from the two semi-metasurfaces. However, some superposition discrepancies appear at the periphery of the flat-top beam. Within the blurred rectangular boxes, the magnitude of the total radiation beam is smaller than the counterpart of beam 2 and even smaller than the counterpart of both beam 1 and beam 2. These discrepancies arise mainly due to the phase inconsistencies of the two quasi-sawtooth-shaped beams across the aperture of the metasurface as discussed in Supplementary Note 4. Fig. 3b plots the phase distribution obtained by mirroring the

phase distribution on the subordinate semi-metasurface to generate a flat-top beam with $\theta_{rf}$ = 50°
and $\theta_p$ = -50°. Its corresponding 3D flat-top radiation beam is illustrated in Fig. 3c, a robust flat-top
radiation beam in the $\varphi = 0°$ cut plane while a narrow beamwidth in the $\varphi = 90°$ can be observed.

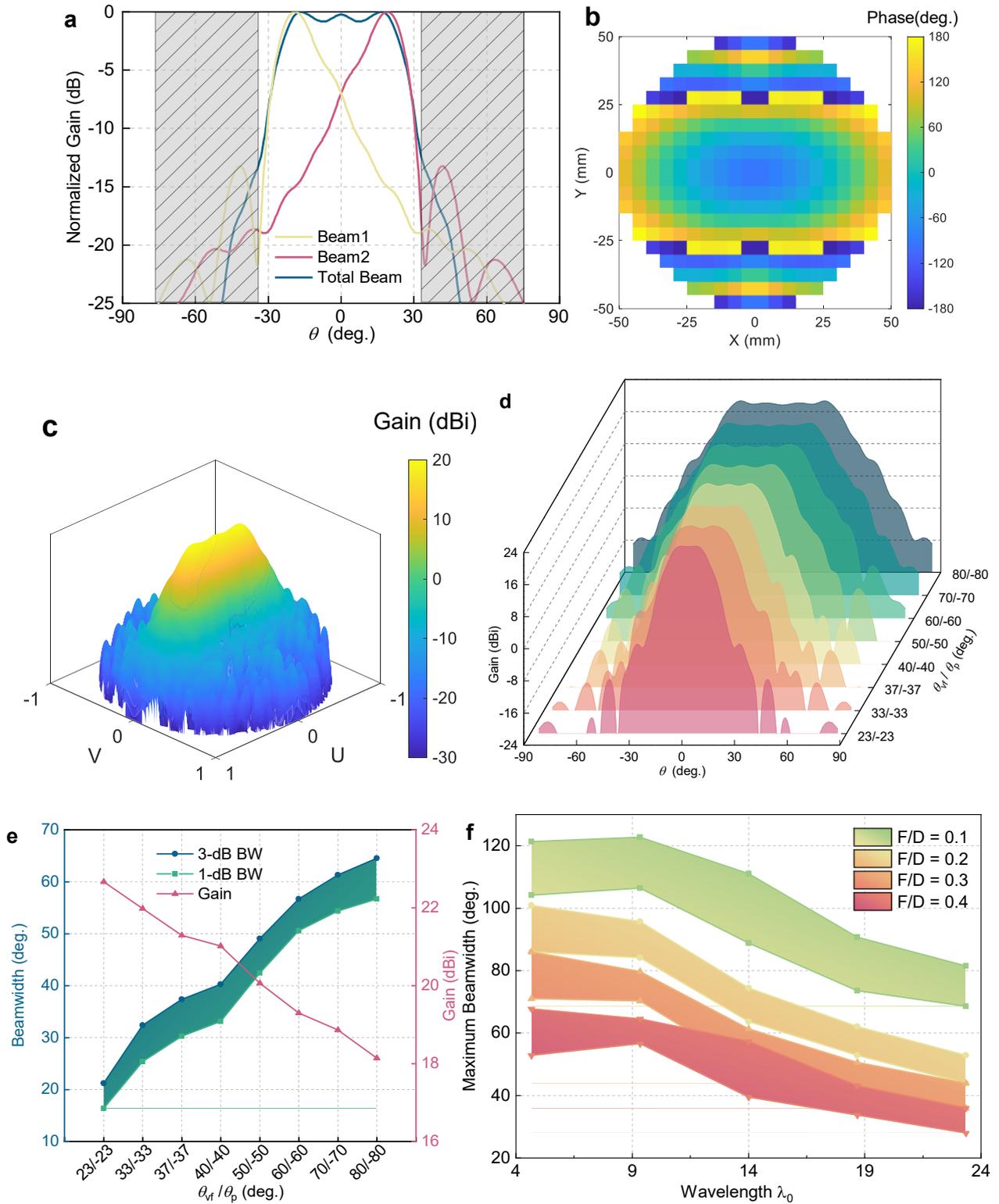

**Fig. 3 | Generation of one-dimensional flat-top beam. a.** The two flipped quasi-sawtooth-shaped beams (red and yellow lines) from the two semi-metasurface and the flat-top beam (dark blue line) from the complete metasurface after performing phase symmetry operations, where $\theta_{vf}$ =50° and $\theta_p$ = -50°. The flat-top beam is quasi-superposed by the two flipped quasi-sawtooth-shaped beams. **b.** The phase distribution to generate a flat-top beam with $\theta_{vf}$ =50° and $\theta_p$ = -50°. **c.** A perspective view of the flat-top beam on the U-V domain with $\theta_{vf}$ =50° and $\theta_p$ = -50°. **d.** The beamwidth of the flat-top beam when ($\theta_{vf}$/ $\theta_p$) increases from 23°/-23° to 80°/-80°. **e.** The relationship between 1-dB, 3-dB beamwidth, gain with ($\theta_{vf}$, $\theta_p$). **f.** The achievable maximum beamwidth (MBW) concerning array aperture size D (in wavelength) and focal distance to aperture diameter ratio F/D. The upper and lower boundaries of the shadow regions are the 3-dB and 1-dB beamwidth, respectively.

As previously concluded, the beamwidth of the superposed flat-top beam is closely associated with the angle pair of ($\theta_{vf}$, $\theta_p$). Here, several flat-top beams with robust flatness within the flat-top range and different beamwidth are computed and displayed under different pairs of ($\theta_{vf}$, $\theta_p$) as depicted in Fig. 3d, where the selection procedure of the angle pairs of ($\theta_{vf}$, $\theta_p$) is detailed in Supplementary Note 5. The 1-dB and 3-dB beamwidths of the flat-top beams range from 16° to 57° and 21° to 65°, respectively. Fig. 3e illustrates the relationships between the beamwidth as well as the gain of the flat-top beam with various pairs of ($\theta_{vf}$, $\theta_p$). The shadowed regions are refined by the upper boundary of the 3-dB bandwidth and the lower boundary of the 1-dB beamwidth. It is anticipated that the gain of the flat-top beam is reduced as both 1-dB and 3-dB beamwidths increase. As seen in Fig. 3d, the beamwidth of the flat-top beam widens as both $\theta_{vf}$ and $\theta_p$ increase, promoting a question here: Is there an upper limit of the beamwidth for the flat-top beam given a specific configuration (the size of the metasurface, the radiation pattern of the external source, and the distance from the external source to the metasurface are all known)? To determine the upper limit, different aperture sizes (D) and focal distance to aperture diameter ratios (F/D) are studied. The $q_f$ = 2 modeling the radiation pattern of the external source at 28 GHz is consistently applied. The aperture size varies from 50 mm to 250 mm (from 4.67 $\lambda_0$ to 23.35 $\lambda_0$ at 28 GHz) with an interval of 50 mm, and the F/D ranges from 0.1 to 0.4 with an interval of 0.1 for each aperture size. The achievable maximum 1-dB and 3-dB beamwidths (MBW) for all these cases are eventually obtained by calculating and comparing the flat-top beam with different angle pairs of ($\theta_{\text{rf}}$, $\theta_{\text{p}}$) and then plotted in Fig. 3f. The detailed angle pairs of ($\theta_{\text{rf}}$, $\theta_{\text{p}}$) are listed in Supplementary Table S1. Since the illumination and spillover efficiencies to the metasurface always remain constant for different aperture sizes when F/D is fixed, the gain of the metasurface reasonably increases with larger aperture size. Therefore, it is anticipated that both 1-dB and 3-dB MBWs are gradually reduced as the aperture size increases for different fixed F/D, aligning well with Fig. 3f. When the aperture size remains constant and F/D is increased, a significant reduction in beamwidth for the achievable 1-dB and 3-dB MBWs is observed accordingly, equivalent to that the corresponding metasurface results in a flat-top beam with a higher gain for a larger F/D. This beamwidth reduction is attributed to the alterations of illumination and spillover efficiencies to the metasurface with different F/D as detailed in Supplementary Note 6. A larger F/D leads to a higher illumination efficiency while the spillover efficiency only drops a little (i.e., from 99% to 93% if F/D is increased from 0.1 to 0.4), ultimately leading to an enhanced aperture efficiency of the metasurface. The enhanced aperture efficiency directly results in a higher gain and narrower beamwidth for the metasurface.

**Flat-top and Isoflux Beams in a Solid Angle**

The preceding findings have effectively demonstrated the ability to generate a flat-top beam in a $\varphi$-cut plane through the implementation of phase symmetry operations that involve semi-metasurface symmetry. Nevertheless, specific applications necessitate flat-top characteristics within all the $\varphi$-cut planes (i.e., within a certain solid angle) to facilitate a broad coverage for two-dimensional wireless communications with a high density of users as illustrated in Fig. 1. Moreover, in some specific applications such as space-to-ground satellite communications where the Earth's geometrically spherical shape must be considered as also illustrated in Fig. 1, isoflux beams featuring a magnitude drop at the boresight direction are highly desirable to offer wider coverage and secure robust link budgets. Achieving a flat-top or isoflux beam within a solid angle is made possible by extending the semi-metasurface phase symmetry operations to rotational phase symmetry operations. This concept transforms the previously defined dominant and subordinate areas in semi-metasurface symmetry operations into dominant and subordinate sequences, which represent a sequence of meta-atoms distributed along the radius, as depicted in Fig. 4a. The phase distribution on the complete metasurface is thus achieved by rotating the subordinate sequence concerning the center of the metasurface, resulting in a rotationally symmetric configuration. The quasi-sawtooth-shaped beams within all $\varphi$-cut planes are expected to be superposed to yield a flat-top or isoflux beam within a solid angle, i.e., 2D flat-top or 2D isoflux beam. The properties of the 2D flat-top or isoflux beam can still be regulated by the same parameters used in manipulating the 1D flat-top or isoflux beam: the tilted angle of the virtual focus $\theta_{vf}$ and the additional phase gradient $\theta_p$. To demonstrate the wide beamwidth of the 2D flat-top beam and wide coverage of the 2D isoflux beam, the aperture size is the same as metasurface studied earlier (i.e., the metasurface consists of 316 meta-atoms forming a circular aperture with the diameter of 20 meta-atoms, and the periodicity of the unit cell is 5 mm) while the F/D is selected as 0.1 based on the findings depicted in Fig. 3f. We determine two sets of three different angle pairs of ($\theta_{vf}$, $\theta_p$) to enable robust 2D flat-top beam and 2D isoflux beam with different beamwidths, respectively. The radiation beams originating from the subordinate sequences are calculated with Eqs. (1) and (3) with two sets of different angle pairs of ($\theta_{vf}$, $\theta_p$) by setting the $|T_{mn}|$ of all meta-atoms excluding the subordinate sequence to be 0. As can be seen in Fig. 4b, all the radiation beams originating from the subordinate sequences with different angle pairs of ($\theta_{vf}$, $\theta_p$) exhibit quasi-sawtooth shapes, which are essential for the generation of a flat-top or isoflux beam. It is also evident that the magnitude of the intersection point at the boresight direction (i.e., $\theta = 0$) for the quasi-sawtooth-shaped beam resulting in a 2D isoflux beam is always less than the counterpart of the quasi-sawtooth-shaped beam resulting in a 2D flat-top beam. By performing rotational phase symmetry operations, the phase distributions to generate 2D flat-top beams with various beamwidths under different pairs of ($\theta_{vf}$, $\theta_p$) are plotted and illustrated in Fig. 4c. It is observed that the phase distributions are all rotationally symmetric. Based on the phase distributions, the radiation beams of the metasurfaces with different pairs of ($\theta_{vf}$, $\theta_p$) are calculated and depicted in Fig. 4c. The highly symmetric 2D flat-top beams with different solid angles for different angle pairs of ($\theta_{vf}$, $\theta_p$) are obtained, where the beamwidth can be tuned from 17° to 100° while still maintaining robust flatness within the flat-top range. Similarly, the phase distributions to generate 2D isoflux beams with various beamwidths under different pairs of ($\theta_{vf}$, $\theta_p$) are plotted and illustrated in Fig. 4d, where rotationally symmetric phase distributions are observed. Then, the corresponding isoflux beams are computed and illustrated in Fig. 4d with the corresponding angle pair of ($\theta_{vf}$, $\theta_p$), where highly symmetric isoflux beams with different beamwidths are obtained. The results illustrated in Fig. 4c, d further validate the efficacy of the phase symmetry operations in manipulating beams in two dimensional.

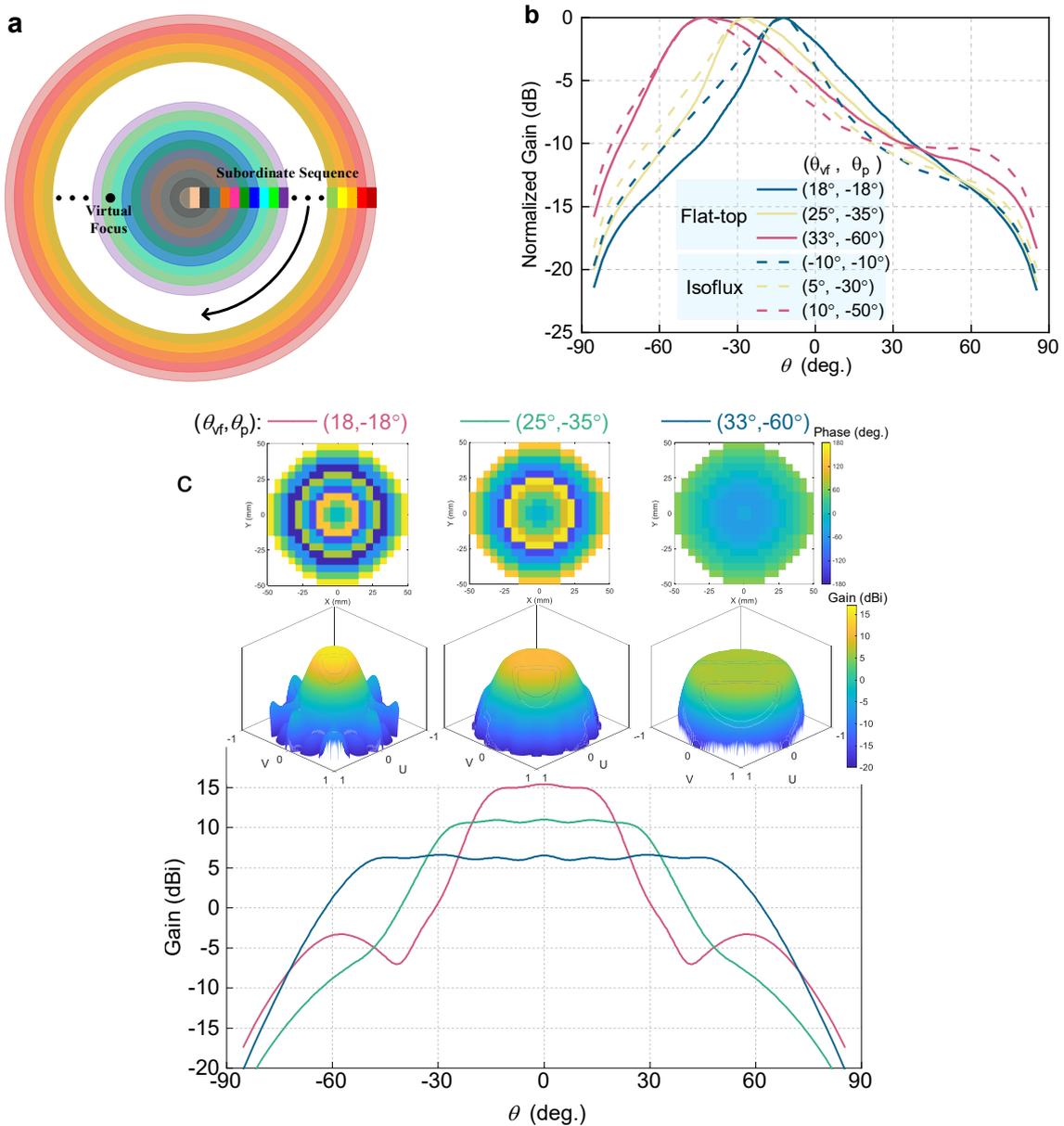

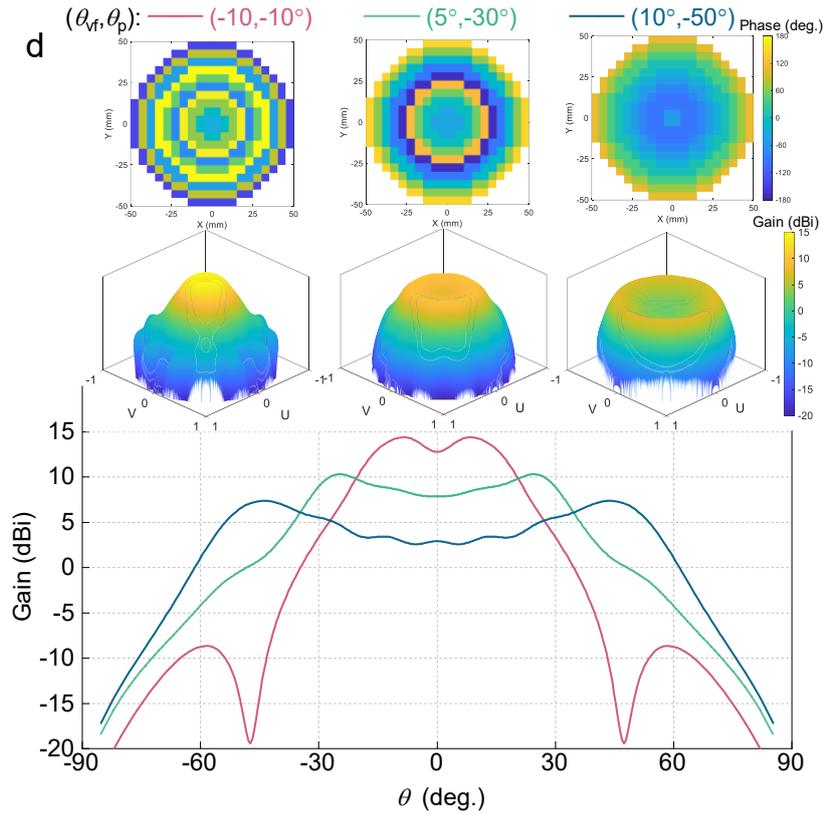

**Fig. 4 | 2D flat-top and isoflux beams by conducting rotational phase symmetry operations. a.** The diagram of performing rotational phase symmetry operations. **b.** The quasi-sawtooth-shaped beams originating from the subordinate sequence with different angle pairs of ($\theta_{vf}$, $\theta_p$) for generating 2D flat-top and isoflux beams. **c.** Three 2D flat-top beams with different beamwidths. **d.** Three isoflux beams with different gain drop at boresight direction.

## Simulation and Experimental Verification

The above analysis has effectively showcased the effectiveness of the phase symmetry operations on efficient beam manipulation for modulated metasurfaces. To further validate this approach, we implement a metasurface comprising four-layered and double metallic rings meta-atoms at 28 GHz, where the frequency response of the meta-atom is detailed in Supplementary Note 7. The metasurface consists of 316 meta-atoms, equivalent to a circular aperture with a diameter of 20 meta-atoms. A linear polarized horn antenna is adopted to centrally illuminate the metasurface with an F/D of 0.4, where the radiation pattern of the horn antenna at 28 GHz aligns closely with the theoretical analysis described earlier and can be approximately modeled with $\cos^{q_f}\theta$ with $q_f$ = 2. With this specific configuration and for proof of the concept, we calculate and plot phase distributions corresponding to one-dimensional flat-top beams with 3-dB beamwidths of 21°, 40°, and 65°, as illustrated in Fig. 5a-c, respectively. Notably, as the beamwidth of the flat-top beam increases, the variation in phase distribution across the aperture of the metasurface becomes progressively smaller along the axis aligned with the flat-top beam (the x-axis in this case). This observation is further supported by the phase distribution on the xoz-plane as illustrated in Fig. 5g-h, where the phase of the electromagnetic wave propagating through the metasurface with a narrow beamwidth undergoes more pronounced disruption in comparison to the metasurface with a wide beamwidth. Based on the phase distributions depicted in Fig. 5a-c, we implement and simulate the corresponding metasurfaces. The simulated radiation beams of these metasurfaces are depicted in the U-V plane as illustrated in Fig. 5d-f, revealing the presence of robust flat-top beams in the $\varphi = 0$ cut plane.

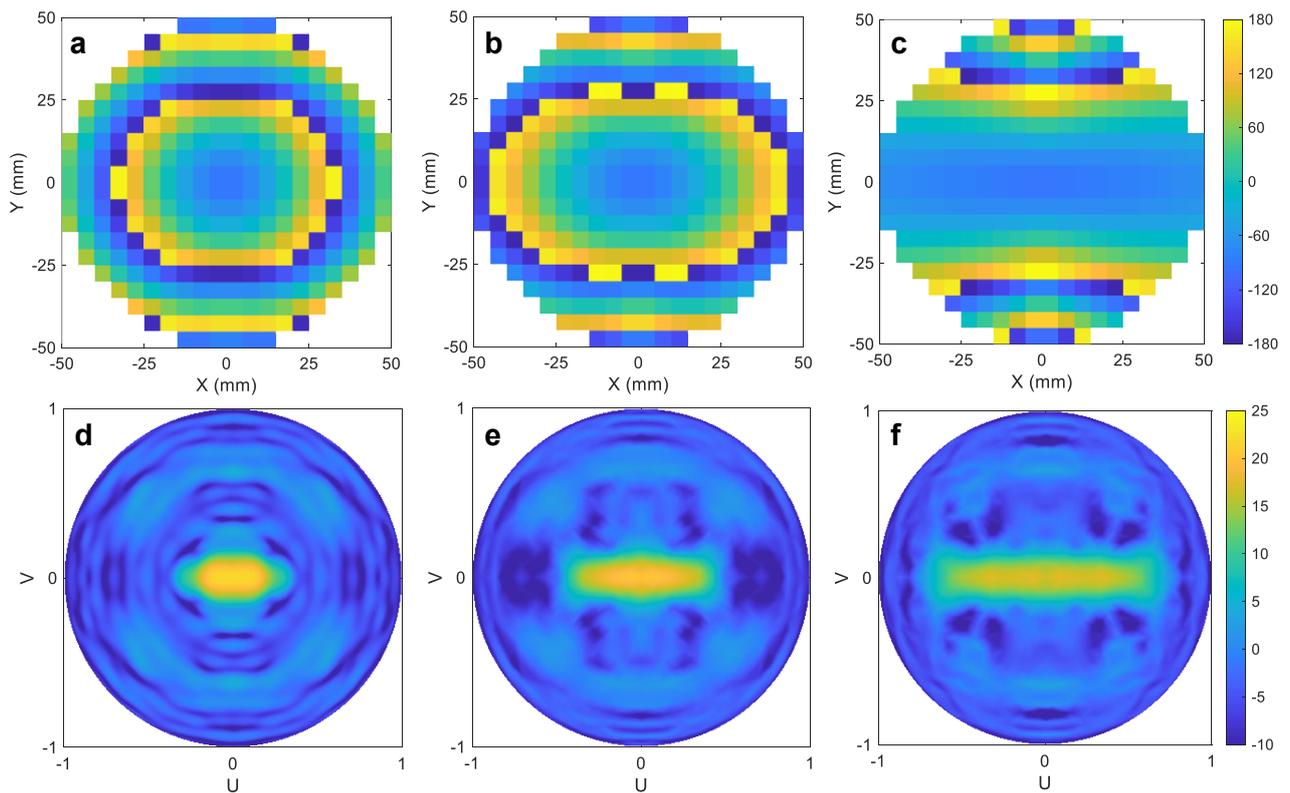

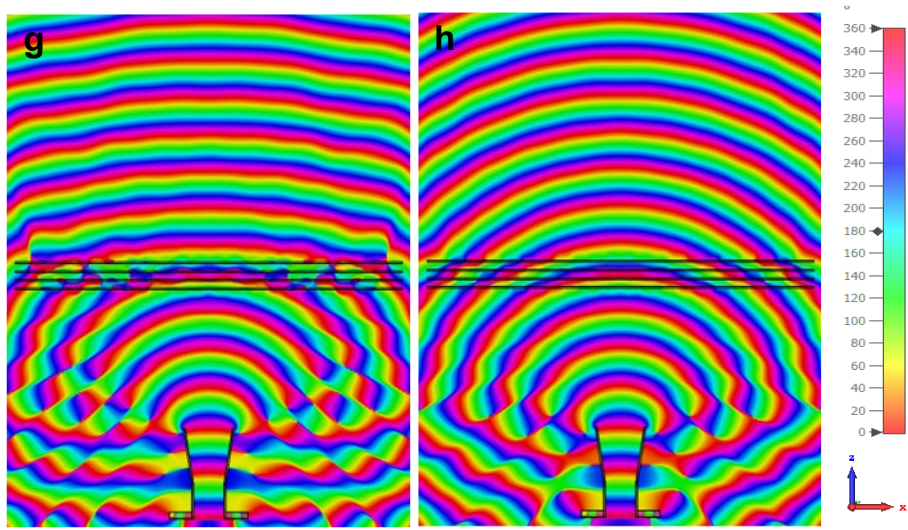

**Fig. 5 | Phase distribution and simulated radiation patterns of three metasurfaces with different flat-top beamwidth. a-c.** Phase distributions for flat-top beams with the 3dB-beamwidth of 21°, 40°, and 65°, respectively. **d-f.** Simulated radiation patterns of these three metasurfaces in the U-V plane. **g.** The phase distribution on the xoz-plane of the metasurface with a narrow flat-top beamwidth. **h.** The phase distribution on the xoz-plane of the metasurface with a wide flat-top beamwidth.

To validate the full-wave simulations, we fabricate and measure two prototypes of the metasurfaces, corresponding to the smallest (Fig. 5d) and largest (Fig. 5f) beamwidths of the one-dimensional flat-top beams. To secure precise alignment between the horn antenna and the metasurface, a customized 3D printing fixture is manufactured to securely assemble them as illustrated in Fig. 6a. Measurements of radiation beam for both metasurfaces are conducted in an anechoic chamber as depicted in Fig. 6a. The measured radiation beams are depicted in the U-V plane as illustrated in Fig. 6b-g at three different frequencies: 27 GHz, 28 GHz, and 29 GHz, clearly demonstrating the presence of flat-top beams for the metasurfaces with narrow and wide beamwidths. To further examine the features of the flat-top beam, the radiation patterns for the metasurfaces with narrow and wide beamwidths at 28 GHz in the $\varphi = 0$ cut plane are displayed in Fig. 6h, i. Additionally, theoretical and simulated results are also provided for comparison. It is evident that the measured results closely match the simulated and theoretical results, providing solid validation of the proposed phase symmetry operations for efficient beam manipulation.

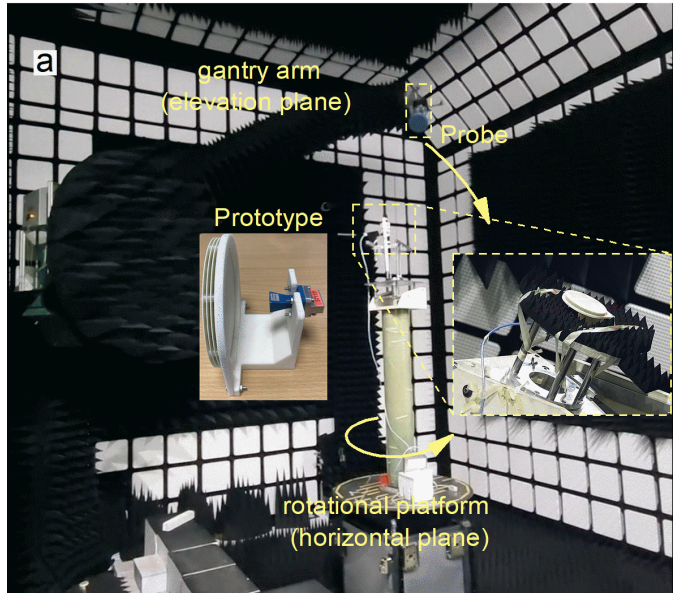
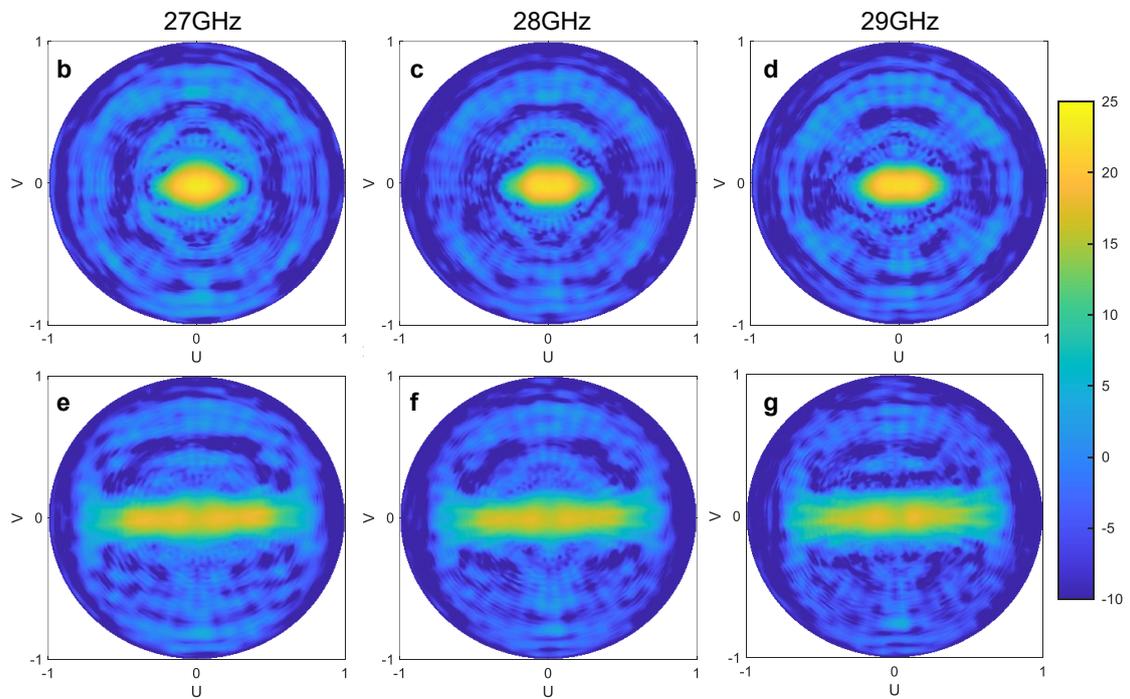
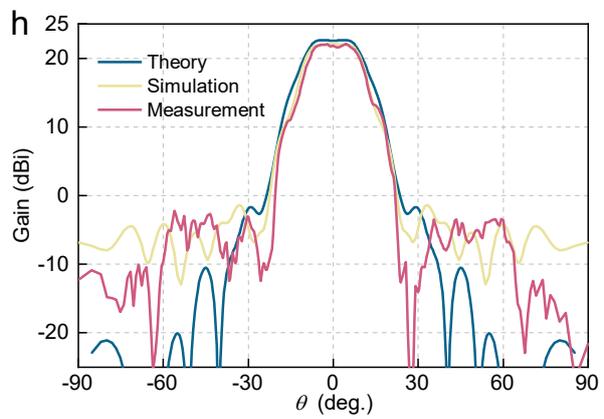
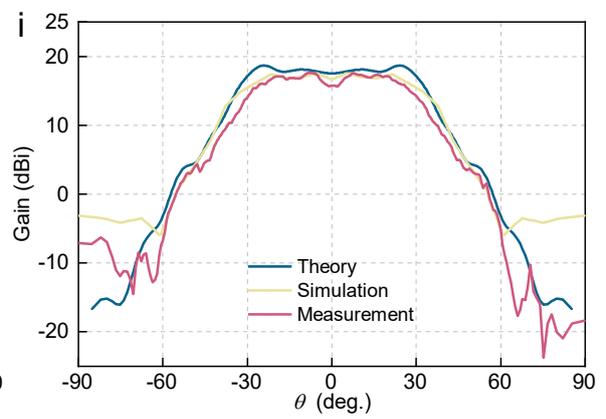

**Fig. 6 | Prototypes and measurement results. a** Measurement setup in an anechoic chamber. **b-g** Measured radiation beam in the upper sphere in the U-V coordinates at 27 GHz, 28 GHz, and 29 GHz. **h, i** Comparison of the radiation beams of theory, simulation, and measurement in the φ = 0 cut plane at 28 GHz.

**Discussion**

In summary, our study introduced an efficient approach for beam manipulation using phase symmetry operations on modulated metasurfaces, with particular emphasis on addressing flat-top and isoflux beams which are challenging to achieve using traditional synthesis methods. By starting from the first principle of field superposition, we identified the fundamental element for generating flat-top and isoflux beams, which is the quasi-sawtooth-shaped beam. The beam can be created by considering only the subordinate semi-surface of the modulated metasurface established through an offset virtual focus ($\theta_{vf}$). By performing phase symmetry operations which involve mirroring the subordinate semi-metasurface to create a complete one, two flipped quasi-sawtooth-shaped beams are obtained, resulting in a superposed beam. Moreover, the quasi-sawtooth-shaped beam can be shifted by adding phase gradient ($\theta_p$) to shape the superposed beam flat-top or isoflux, depending on the magnitude of the quasi-sawtooth-shaped beam at the boresight direction compared to its peak magnitude. Similarly, 2D flat-top or isoflux beams can be realized by performing rotational phase symmetry operations which involve rotating the subordinate sequence concerning the center of the metasurface. Remarkably, the beamwidths of the flat-top or isoflux beams can be flexibly controlled by selecting appropriate angle pairs ($\theta_{vf}$, $\theta_p$) often with minimal optimization required. Experimental measurements have validated the effectiveness of the phase symmetry operations in generating flat-top beams with adjustable beamwidths. The concept of phase symmetry operations can be extended to various vectorial components, potentially offering new avenues for manipulating different types of waves including acoustic waves, water surface waves, and beyond to advance related applications.

**Method**

**Numerical calculations and simulations**

The radiation pattern was calculated using the array theory, and the fast Fourier transform (FFT) was utilized to accelerate the computing[41,42]. The simulations were performed in the commercial software CST Microwave Studio.

**Details on the prototype**

The prototypes of the metasurfaces were fabricated using conventional printed circuit board (PCB) technology. Plastic pads, each with a thickness of 2 mm, are manufactured with CNC technology to precisely separate the layers of the prototype. The layers were assembled with plastic screws.

**Experimental setup**

The measurement of the far-field radiation beam of the prototypes was conducted in an anechoic chamber at Aalborg University. The prototype was affixed to a rotational platform capable of horizontal plane rotation. The gantry arm is 3 meters away from the metasurface and moves in the elevation plane. The scan covered a 360° range with a step of 5° in the horizontal plane and 90° range with a step of 1° in the elevation plane, respectively. The external source used in this article is a commercial horn antenna with a model of "PASTERNACK PE9851/2F-10". Dual linear polarization measurements were conducted to transform into co-, and cross- polarization.

## Acknowledgments

This work was partially supported by the China Scholarship Council (CSC).

## Author contributions

Y. C. and P. M. contributed equally to this work. P. M. conceived the concept. Y. C. and P. M. carried out the derivation and analysis of the phase symmetry operations. Y. C. conducted the simulation and measurement of the modulated metasurfaces with flat-top and isoflux beams. G. F. P. and S. Z. supervised this project. All the authors contribute to writing the manuscript.

## Competing interests

The authors declare no competing interests.

# Efficient Beam Manipulation with Phase Symmetry Operations on Modulated Metasurfaces

*Yang Cai, Peng Mei\*, Gert Frølund Pedersen, Shuai Zhang*

Antennas, Propagation and Millimeter-wave Systems (APMS) section, Department of Electronic Systems, Aalborg University, Denmark

Corresponding author: *Peng Mei*, mei@es.aau.dk

This supplementary information contains the following sections:

**Supplementary Note 1: Mathematical principles to generate flat-top and isoflux shapes.**

**Supplementary Note 2: Analysis and comparison of radiation beams originating from the dominant and subordinate semi-surfaces.**

**Supplementary Note 3: Generation of one-dimensional isoflux beam.**

**Supplementary Note 4: Field superpositions from two flipped quasi-sawtooth-shaped beams to a superposed beam.**

**Supplementary Note 5: Derivation of the optimal parameters for robust flat-top beams.**

**Supplementary Note 6: The effects of the aperture size and F/D on the maximum achievable beamwidths of flat-top beams.**

**Supplementary Note 7: Meta-atom analysis.**

**Supplementary Note 1: Mathematical principles to generate flat-top and isoflux shapes.**

Assuming we have a beam that can be formulated as a shifted ramp function (or can refer to as a sawtooth-shaped beam):

$$y = \begin{cases} a(\theta - \theta_2), & (\theta_1 < \theta < \theta_2) \\ 0, & \text{others} \end{cases} \quad (S1)$$

where $a$ is a coefficient representing the slope of the ramp. We shift the beam by $\alpha_p$ so that the magnitude at $\theta = 0$ is half of the maximal value as shown in Supplementary Fig. S1a, and the shift beam and angle $\alpha_p$ are

$$y = \begin{cases} a(\theta - \theta_2 - \alpha_p), & (\theta_1 + \alpha_p < \theta < \theta_2 + \alpha_p) \\ 0, & \text{others} \end{cases} \quad (S2)$$

$$\alpha_p = -\frac{\theta_1 + \theta_2}{2} \quad (S3)$$

A mirrored beam concerning the shift beam expressed in Eq. (S2) can be formulated as:

$$y = \begin{cases} a(-\theta - \theta_2 - \alpha_p), & (-\theta_2 - \alpha_p < \theta < -\theta_1 - \alpha_p) \\ 0, & \text{others} \end{cases} \quad (S4)$$

If we superpose the shifted beam to the mirrored one, the sum beam would be:

$$\begin{aligned} y &= \begin{cases} a(\theta - \theta_2 - \alpha_p) + a(-\theta - \theta_2 - \alpha_p), & (-\theta_t < \theta < \theta_t) \\ 0, & \text{others} \end{cases} \\ &= \begin{cases} 2a(-\theta_2 - \alpha_p), & (\theta_1 + \alpha_p < \theta < \theta_2 + \alpha_p) \\ 0, & \text{others} \end{cases} \\ &= \begin{cases} a(\theta_1 - \theta_2), & (\theta_1 + \alpha_p < \theta < \theta_2 + \alpha_p) \\ 0, & \text{others} \end{cases} \end{aligned} \quad (S5)$$

which is a square wave as illustrated in Supplementary Fig. S1b, and the magnitude remains constant within the range of $[(\theta_1 - \theta_2)/2, (\theta_2 - \theta_1)/2]$ after substituting Eq. (S3), equivalent to a flat-top beam. If we shift the original beam in Eq. (S1) further so that the magnitude at $\theta = 0$ is smaller than half of the maximal value as depicted in Supplementary Fig. S1c, the sum beam would be a concave shape, equivalent to an isoflux beam.

$$y = \begin{cases} a(\theta - \theta_2 - \alpha_p), & (\theta_1 + \alpha_p < \theta < -\theta_2 - \alpha_p) \\ 2a(-\theta_2 + \alpha_p), & (-\theta_2 - \alpha_p < \theta < \theta_2 + \alpha_p) \\ a(-\theta - \theta_2 - \alpha_p), & (\theta_2 + \alpha_p < \theta < -\theta_1 - \alpha_p) \\ 0, & \text{others} \end{cases} \quad (S6)$$

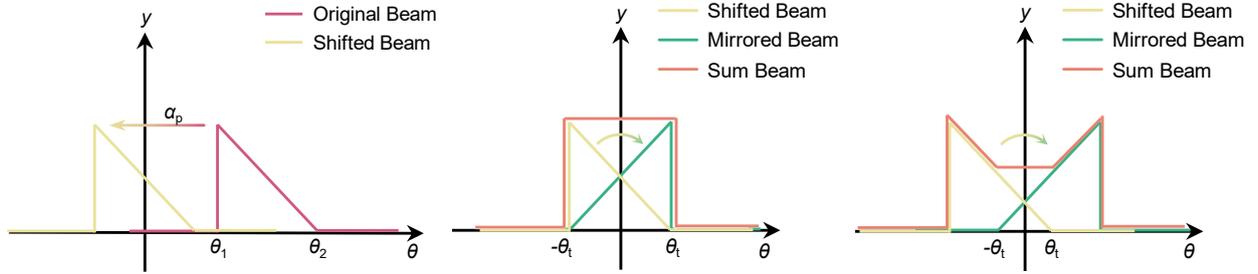

**Supplementary Fig. S1 | Mathematical principles. a.** The original beam shifts along the $\theta$-axis with an angle of $\alpha_p$. **b.** When the shift angle $\alpha_p = (\theta_1 + \theta_2)/2$, two ramp beams will be superposed into a square beam. **c.** When the shift angle $\alpha_p > (\theta_1 + \theta_2)/2$, two ramp beams will be superposed into a concave shape.

**Supplementary Note 2: Analysis and comparison of radiation beams originating from the dominant and subordinate semi-surfaces.**

This supplementary note clarifies the reasons for determining the subordinate semi-surface to perform phase symmetry operations, which is also applicable to the subordinate semi-sequence in the generation of 2D flat-top and isoflux beams. In principle, we can utilize either the dominant semi-surface or subordinate semi-surface to conduct phase symmetry operations. Here, we calculate the radiation beams originating from the dominant and subordinate semi-surfaces with Eq. (1) with different tilted angles of the virtual focus $\theta_{vf}$. As shown in Supplementary Fig. S2a, the quasi-sawtooth-shaped beam resulting from the dominant semi-surface deteriorates quickly when the virtual focus tilts to a large angle. In contrast, the quasi-sawtooth-shaped beam originating from the subordinate semi-surface remains in relatively good shape even at a large tilted angle of the virtual focus as illustrated in Supplementary Fig. S2b. It is of paramount importance to keep the quasi-sawtooth shape of the beam since any distortion in the quasi-sawtooth-shaped beam will result in a degraded flat-top beam with pronounced ripples. The subordinate semi-surface is, therefore, the preferred option for the generation of a flat-top beam with a robust flatness.

We investigate the reasons for the distinct quasi-sawtooth-shaped beams originating from the dominant semi-surface and subordinate semi-surface. From the perspective of field superposition, the radiation beam from each semi-surface is the vectorial superposition of electric fields (i.e., amplitude and phase) from all meta-atoms that each semi-surface includes. It should be noted here when we calculate the radiation beams originating from the two semi-surfaces, the amplitude distributions across the two semi-surfaces are the same since the external source is located on-axis. The main difference is the phase distributions across the two semi-surfaces, which is the key factor responsible for generating distinct radiation beams. Here, we compute and depict the phase distributions across the two semi-surfaces with different tilted angles of the virtual focus ($\theta_{vf}$) for an on-axis external source, where the free-space phase delay from the on-axis external source to the two semi-surfaces is also considered, as illustrated in Supplementary Fig. S2c. The gradients of the phase distributions across the two semi-surfaces at different tilted angles of the virtual focus ($\theta_{vf}$) are also calculated and illustrated in Supplementary Fig. S2d. It is evident that the phase variations on the subordinate semi-surface are smaller than the counterparts on the dominant semi-surface, especially much smaller when the tilted angle of the virtual focus ($\theta_{vf}$) reaches a large value. The difference in phase gradient results in different scanned angles and radiation patterns of the radiation beams originating from the two semi-surfaces for a large tilted angle of the virtual focus ($\theta_{vf}$).

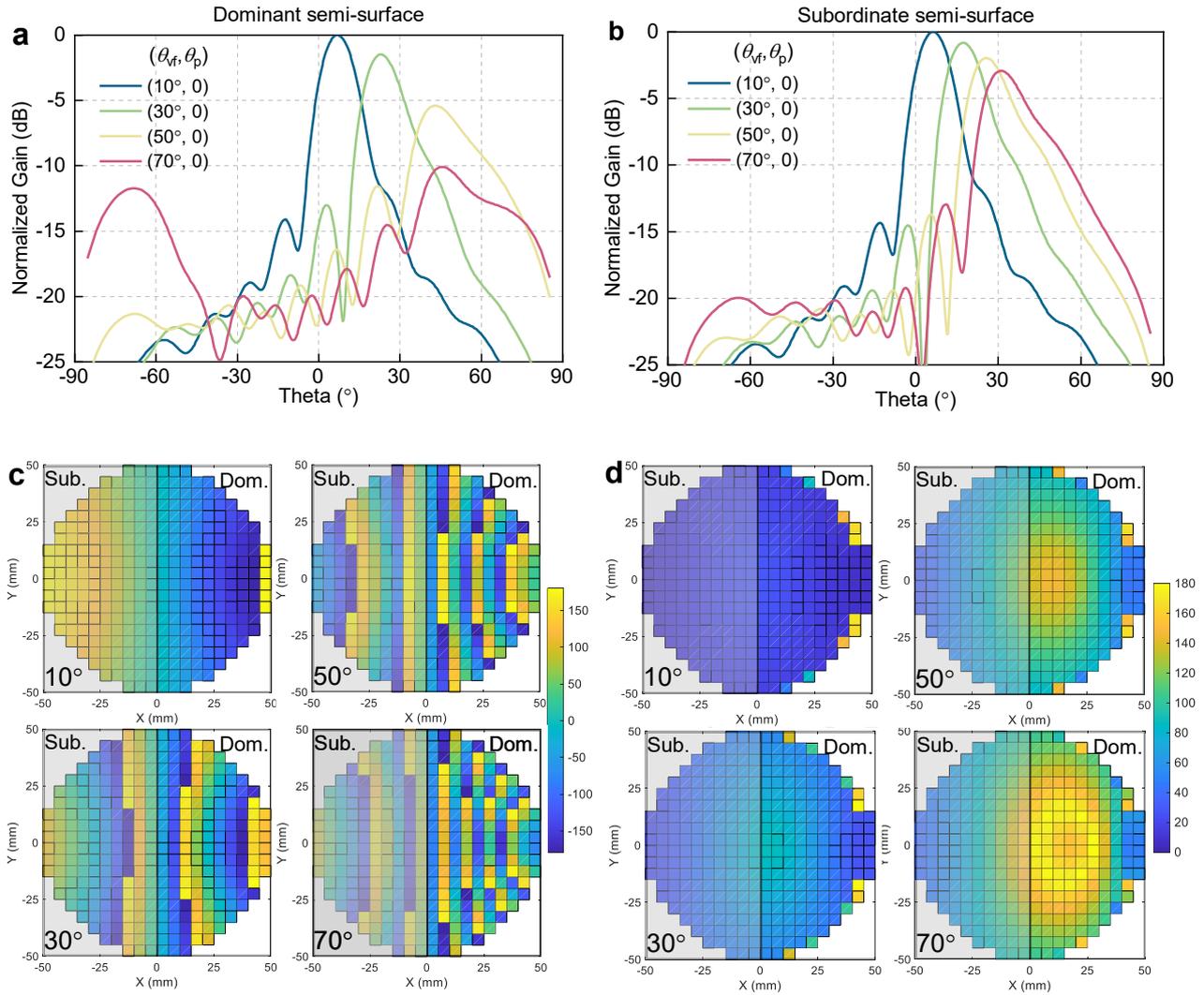

**Supplementary Fig. S2 | Analysis and comparison of radiation beams originating from the dominant and subordinate semi-surfaces. a.** The radiation beam originating from the dominant semi-surface with different $\theta_{vf}$. **b.** The radiation beam originating from the subordinate semi-surface with different $\theta_{vf}$. **c.** The phase distributions across the two semi-surfaces with different titled angles of the virtual focus ($\theta_{vf}$) for the on-axis external source, where phase delay from the on-axis external source to the two semi-surfaces is also considered. **d.** The gradient of the phase distributions across the two semi-surfaces with different titled angles of the virtual focus ($\theta_{vf}$)

**Supplementary Note 3: Generation of a one-dimensional isoflux beam.**

This supplementary note clarifies the generation of a one-dimensional isoflux beam. As discussed in Supplementary Note 1, the magnitude of the quasi-sawtooth-shaped beam at the intersection point (i.e., $\theta = 0$) has a crucial effect on the formation of a flat-top or isoflux beam. A higher magnitude at the intersection point (i.e., $\theta = 0$) will lead to the superposed beam to converge into a pencil beam. Conversely, a lower magnitude at the intersection point (i.e., $\theta = 0$) will facilitate the superposed beam to shape into an isoflux beam. As illustrated in Fig. 2f, the magnitude of the quasi-sawtooth-shaped beam at the intersection point (i.e., $\theta = 0$) can be flexibly shifted by imposing proper additional phase gradient $\theta_p$. Here, we compute and depict the quasi-sawtooth-shaped beam originating from the subordinate semi-surface with different additional phase gradients $\theta_p$ as depicted in Supplementary Fig. S3a, where the titled angle of the virtual focus $\theta_{vf}$ is fixed and equal to 50°. Specifically speaking, the magnitude of the quasi-sawtooth-shaped beam at the intersection point (i.e., $\theta = 0$) is maximal when $\theta_p = -25°$, the magnitude of the quasi-sawtooth-shaped beam at the intersection point (i.e., $\theta = 0$) is 6 dB lower than its peak magnitude when $\theta_p = -50°$, the magnitude of the quasi-sawtooth-shaped beam at the intersection point (i.e., $\theta = 0$) is more than 6 dB lower than its peak magnitude when $\theta_p = -75°$. After performing phase symmetry operations, the corresponding superposed beams are illustrated in Supplementary Fig. S3b, where pencil, flat-top, and isoflux beams are all achieved and observed at respective additional phase gradient $\theta_p$.

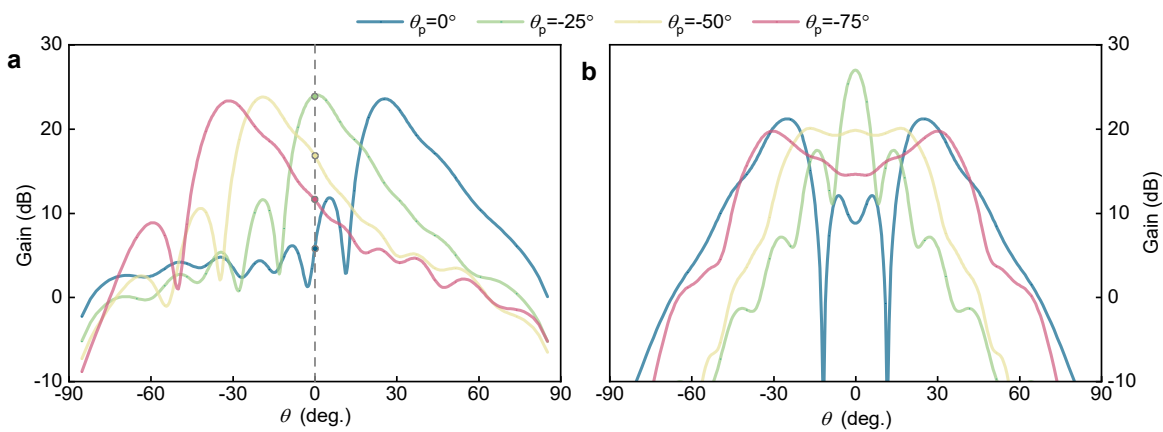

**Supplementary Fig. S3 | Radiation beam originating from the subordinate semi-surface and the complete surface after performing phase symmetry operations. a.** The quasi-sawtooth-shaped beam generated by the subordinate semi-surface with fixed $\theta_{vf} = 50°$ is flexibly shifted with different additional phase

gradients $\theta_p$. **b.** The superposed beams originating from the complete surface after performing phase symmetry operations with different additional phase gradients $\theta_p$.

**Supplementary Note 4: Field superpositions from two flipped quasi-sawtooth-shaped beams to a superposed beam.**

After performing phase symmetry operations, the superposed beam resulting from the complete surface is the vectorial superposition (i.e., magnitude and phase) of the two flipped quasi-sawtooth-shaped beams originating from the two semi-surfaces. To demonstrate the validity of the strict vectorial superposition, the E-fields originating from the subordinate semi-metasurface, the mirrored semi-metasurface obtained by conducting phase symmetry operations, and the complete metasurface comprising the two semi-surfaces are computed and analyzed. As consistent with the manuscript, the tilted angle of the virtual focus and the additional phase gradient are 50º and -50º, respectively. Using Eqs. (1)-(3), the magnitudes of the E-field resulting from the subordinate semi-metasurface, its mirrored semi-metasurface, and the complete metasurface comprising the two semi-surfaces are computed and depicted in Fig. S4a. Also, the phases of the E-field originating from the subordinate semi-metasurface, its mirrored semi-metasurface, and the complete metasurface comprising the two semi-surfaces are computed and illustrated in Supplementary Fig. S4b. It is evident that the phases of the E-fields from the two semi-metasurfaces are the same within a small range around $\theta = 0º$ denoted with a light blurred box in Supplementary Fig. S4b, which results in direct and precise magnitude superposition for the superposed beam as also denoted with a light blurred box in Fig. S4a. Specifically, the magnitudes of the E-field of the two flipped quasi-sawtooth-shaped beams originating from the two semi-metasurfaces are 683 at $\theta = 0º$, while the magnitude of the E-field of the complete metasurface is 1366 that is exactly twice the value of 683. When $\theta$ becomes large, the phases of the E-field of the two flipped quasi-sawtooth-shaped beams originating from the two semi-metasurfaces deviate gradually as denoted with dark blurred boxes in Supplementary Fig. S4b. The phase deviation explains the results illustrated in Supplementary Fig. S4a that the magnitude of the superposed beam is smaller than the magnitude of the beam from either the subordinate semi-metasurface and its mirrored semi-metasurface, and sometimes is smaller than the magnitudes of the beams from both subordinate semi-metasurface and its mirrored semi-metasurface as denoted with dark blurred boxes in Supplementary Fig. S4a.

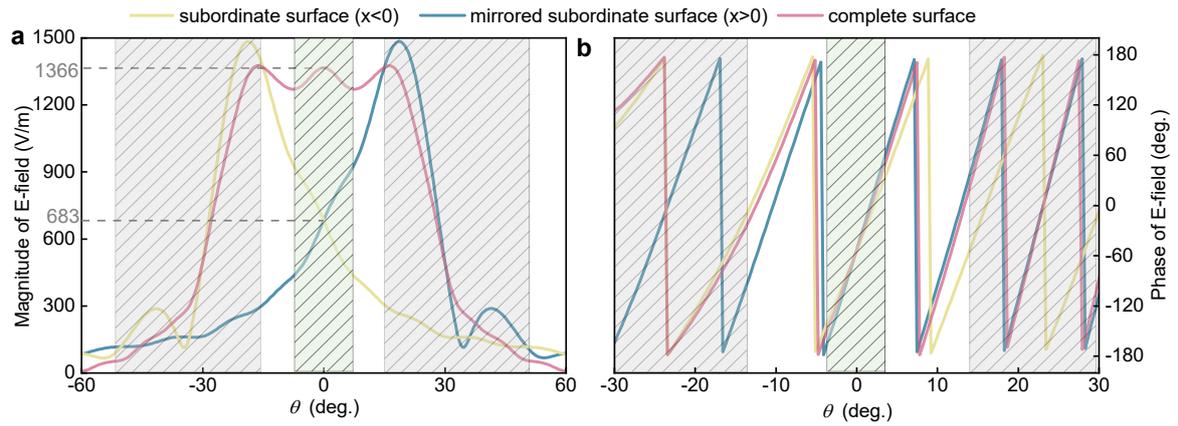

**Supplementary Fig. S4 | Theoretical E-field in a cut plane. a.** The magnitudes of the two flipped quasi-sawtooth-shaped beams originating from the two semi-metasurfaces and the superposed beam of the complete metasurface comprising the two semi-metasurfaces. **b.** The phases of the two flipped quasi-sawtooth-shaped beams originating from the two semi-metasurfaces and the superposed beam of the complete metasurface comprising the two semi-metasurfaces.

**Supplementary Note 5: Derivation of the optimal parameters for robust flat-top beams.**

To synthesize a flat-top beam, the optimal tilted angle of virtual focus $\theta_{vf}$ and additional gradient phase $\theta_p$ can be derived according to Eq. (S3). Firstly, $\theta_{vf}$ is usually determined based on the desired beamwidth. The beamwidth of the synthesized flat-top beam is in positive correlation with the $\theta_{vf}$. Specifically speaking, a large $\theta_{vf}$ usually lead to a wider beamwidth of the flat-top beam. On the other hand, as we can observe from Supplementary Fig. S2b, the 6-dB gain beamwidth of the quasi-sawtooth-shaped beams broadens with $\theta_{vf}$, which also facilitate a wider superposed flat-top beam. When $\theta_{vf}$ is fixed, a quasi-sawtooth-shaped beam originating from the subordinate semi-surface is readily generated, where we can determine the corresponding $\theta_1$ and $\theta_2$ as defined in Supplementary Fig. S1a. Then, the optimal value of $\theta_p$ can be initially calculated by Eq. (S3). Since it is a quasi-sawtooth-shaped beam, it is difficult to determine precise $\theta_1$ and $\theta_2$. The calculated $\theta_p$, therefore, needs further fine-tuning. Alternatively, the optimal value of $\theta_p$ can also be obtained by observing the scan angle of the quasi-sawtooth-shaped beam since $\theta_p$ can flexibly shift the quasi-sawtooth-shaped beam in the angular domain as illustrated in Supplementary Fig.S3a. As a result, we can always determine a proper $\theta_p$ to satisfy the 6dB gain drop at the boresight direction (i.e., $\theta = 0$) for a given $\theta_{vf}$. Ultimately, we can synthesize a flat-top beam with the desired beamwidth, requiring minimal optimization throughout this process.

**Supplementary Note 6: The effects of the aperture size and F/D on the maximum achievable beamwidths of flat-top beams.**

The gain of the metasurface is directly determined by the aperture efficiency of the metasurface, while the aperture efficiency is closely associated with the illumination and spillover efficiencies. Illumination efficiency ($\eta_{ill}$) describes how uniform of the electric field is distributed across the metasurface, while spillover efficiency ($\eta_{spill}$) is defined as the percentage of the radiated power from the external source that is intercepted by the metasurface. The illumination and spillover efficiencies can be evaluated with the following formulae [1]:

$$\eta_{ill} = \frac{1}{A_p}\frac{\left[\int_s I(m,n)dS\right]^2}{\int_s |I(m,n)|^2 dS} = \frac{1}{A_p}\frac{\left|\sum_{m=1}^{M}\sum_{n=1}^{N} I(m,n)\Delta x\Delta y\right|^2}{\sum_{m=1}^{M}\sum_{n=1}^{N}|I(m,n)|^2 \Delta x\Delta y}, \tag{S7}$$

$$\eta_{spill} = \frac{2q_f+1}{2\pi}\sum_{m=1}^{M}\sum_{n=1}^{N}\frac{F\cos^{2q_f}(\theta_f)}{r^3}\Delta x\Delta y, \tag{S8}$$

where $A_p$ is the aperture area of the metasurface ($A_p \cong \pi D^2/4$), $I(m, n)$ is the aperture field as evaluated in (5), $dS$ is the meta-atom area, $q_f$ is the power factor of the illumination source, $F$ is the focal distance, $r$ is the distance between mn[th] meta-atom and the source, $\Delta x$ and $\Delta y$ are the meta-atom size along $x$- and $y$-axes, respectively.

For a lossless metasurface, the aperture efficiency of the metasurface is equal to illumination efficiency multiplying spillover efficiency (i.e., $\eta_{total} = \eta_{ill} \times \eta_{spill}$). When F/D is fixed, the illumination and spillover efficiencies almost remain constant with different aperture size of the metasurface as illustrated in Supplementary Fig. S5a. The gain (i.e., G = D * $\eta_{total}$) of the metasurface, thus, increases with the aperture size D, consequently leading to a narrower MBW. On the other hand, when D is fixed, the illumination edge taper level rises dramatically from -40 dB to -10 dB as F/D increases from 0.1 to 0.4 as illustrated in Supplementary Fig. S5b. The illumination efficiency increases substantially with increment of F/D, while the spillover efficiency just drops a little, finally leading to the aperture efficiency $\eta_{total}$ increases from 0.21 to 0.79 as illustrated in Supplementary Fig. S5a. The enhanced aperture efficiency increases the gain of the metasurface, resulting in a narrower MBW.

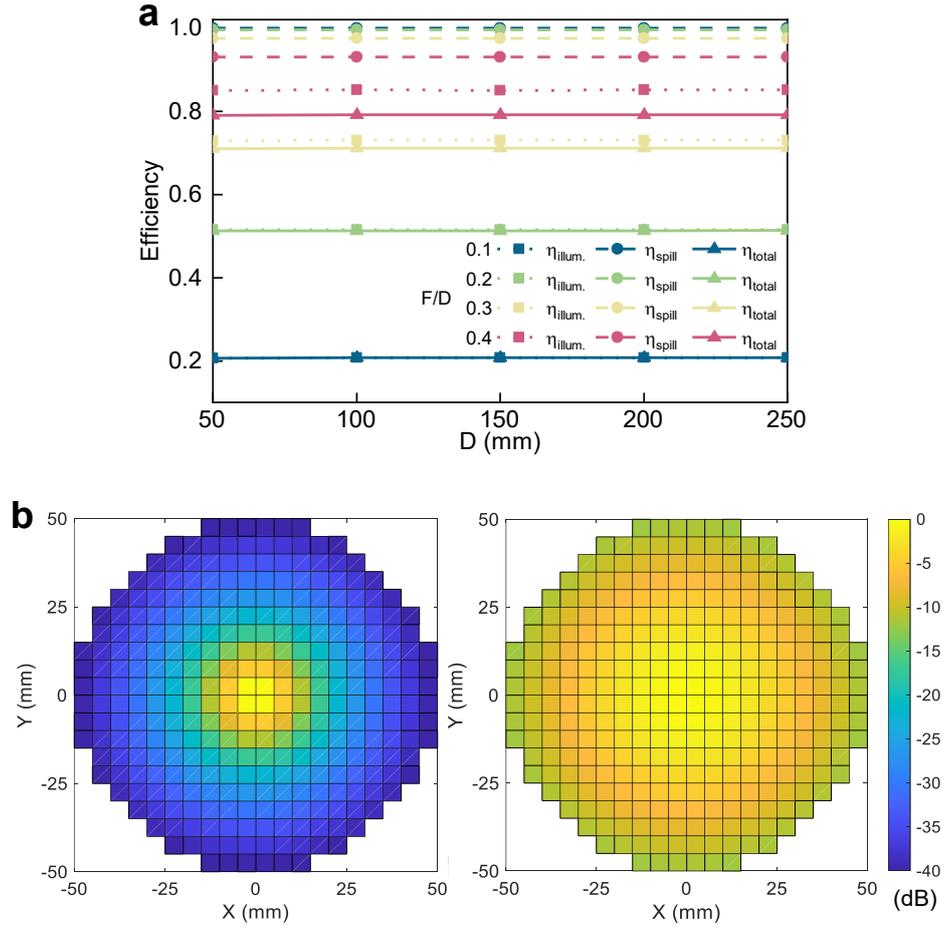

**Supplementary Fig. S5 | Efficiency and amplitude distribution of the metasurfaces. a.** The illumination, spillover, and total efficiency under different configurations of D and F/D. **b**. The amplitude on the metasurfaces when D = 100 mm, F/D =0.1 (left), and F/D = 0.4 (right).

**Supplementary Table. S1| The beamwidth under different configurations of the metasurfaces.**

| F/D | 0.1 | | | | |
|---|---|---|---|---|---|
| **D (mm)** | 50 | 100 | 150 | 200 | 250 |
| **($\theta_{vf}$, $\theta_p$) (deg.)** | (0, 90) | (50, 90) | (60, 90) | (70, 90) | (70, 90) |
| **1-dB MBW (deg.)** | 104 | 106 | 89 | 74 | 69 |
| **3-dB MBW (deg.)** | 121 | 123 | 111 | 91 | 81 |
| **F/D** | 0.2 | | | | |
| **D (mm)** | 50 | 100 | 150 | 200 | 250 |
| **($\theta_{vf}$, $\theta_p$) (deg.)** | (55, 90) | (70, 90) | (75, 90) | (80, 90) | (80, 80) |
| **1-dB MBW (deg.)** | 86 | 84 | 64 | 53 | 44 |
| **3-dB MBW (deg.)** | 101 | 96 | 74 | 62 | 53 |
| **F/D** | 0.3 | | | | |
| **D (mm)** | 50 | 100 | 150 | 200 | 250 |
| **($\theta_{vf}$, $\theta_p$) (deg.)** | (70, 90) | (70, 90) | (80, 80) | (80, 80) | (80, 80) |
| **1-dB MBW (deg.)** | 71 | 70 | 49 | 40 | 36 |
| **3-dB MBW (deg.)** | 86 | 80 | 61 | 51 | 44 |

| F/D | 0.4 | | | | |
|---|---|---|---|---|---|
| **D (mm)** | 50 | 100 | 150 | 200 | 250 |
| **($\theta_{vf}$, $\theta_p$) (deg.)** | (80, 80) | (80, 80) | (90, 80) | (90, 80) | (90, 80) |
| **1-dB MBW (deg.)** | 53 | 57 | 40 | 34 | 30 |
| **3-dB MBW (deg.)** | 68 | 65 | 57 | 43 | 36 |

**Supplementary Note 7: Meta-atom analysis**

Any meta-atom capable of simultaneous high transmission efficiency and full phase-cycle coverage can be adopted to implement the metasurface described in this paper. Without loss of generality, a four-layered double-ring meta-atom is designed and used to build on the metasurface to showcase the flexible beam manipulations enabled by phase symmetry operations [3]. As shown in Supplementary Fig. S6a, the meta-atom consists of four identical layers separated each other with an air spacer, where two double metallic rings are printed on supporting substrate (the substrate used here is Rogers RO4350B with a loss tangent of 0.0027). The equivalent circuit of the single layer is extracted and shown in Supplementary Fig. S6d. Each metallic ring can be modeled with a series of inductor (L) and capacitor (C), and the supporting substrate can be equivalent with a transmission line. It is known that a shunt LC demonstrates a bandstop property at a specific frequency that is calculated by $f = 1/(2\pi\sqrt{LC})$, leading to a transmission zero in the frequency response. Due to the different dimensions of the double metallic rings, the frequencies of the two transmission zeros are distinguished accordingly, eventually resulting in a reflection pole in between. The current distribution on the double metallic rings is simulated and plotted at the frequency of reflection pole as illustrated in Supplementary Fig. S6a, where it is observed that currents are mainly concentrated on the outer periphery of the inner metallic ring and inner periphery of the outer ring. The current distribution indicates that the frequency response of the single layer can be controlled by either tuning the radius of the inner metallic ring or the radius of the outer metallic ring. It has been proved that a single layer double metallic ring can only offer a maximum of 90-degree phase coverage. By cascading the single layer into four layers as shown in Supplementary Fig. S6a, it can demonstrate simultaneous high transmission efficiency and 360-degree phase coverage. The transmission amplitude and phase of the meta-atom are illustrated in Supplementary Fig. S6e with different values of $r_2$ at 28GHz, where it is observed that the meta-atom exhibits over 360-degree phase coverage when the value of $r_2$ is varied from 0.85 to 1.35 mm and the attenuation is smaller than 1.5 dB within the phase coverage simultaneously.

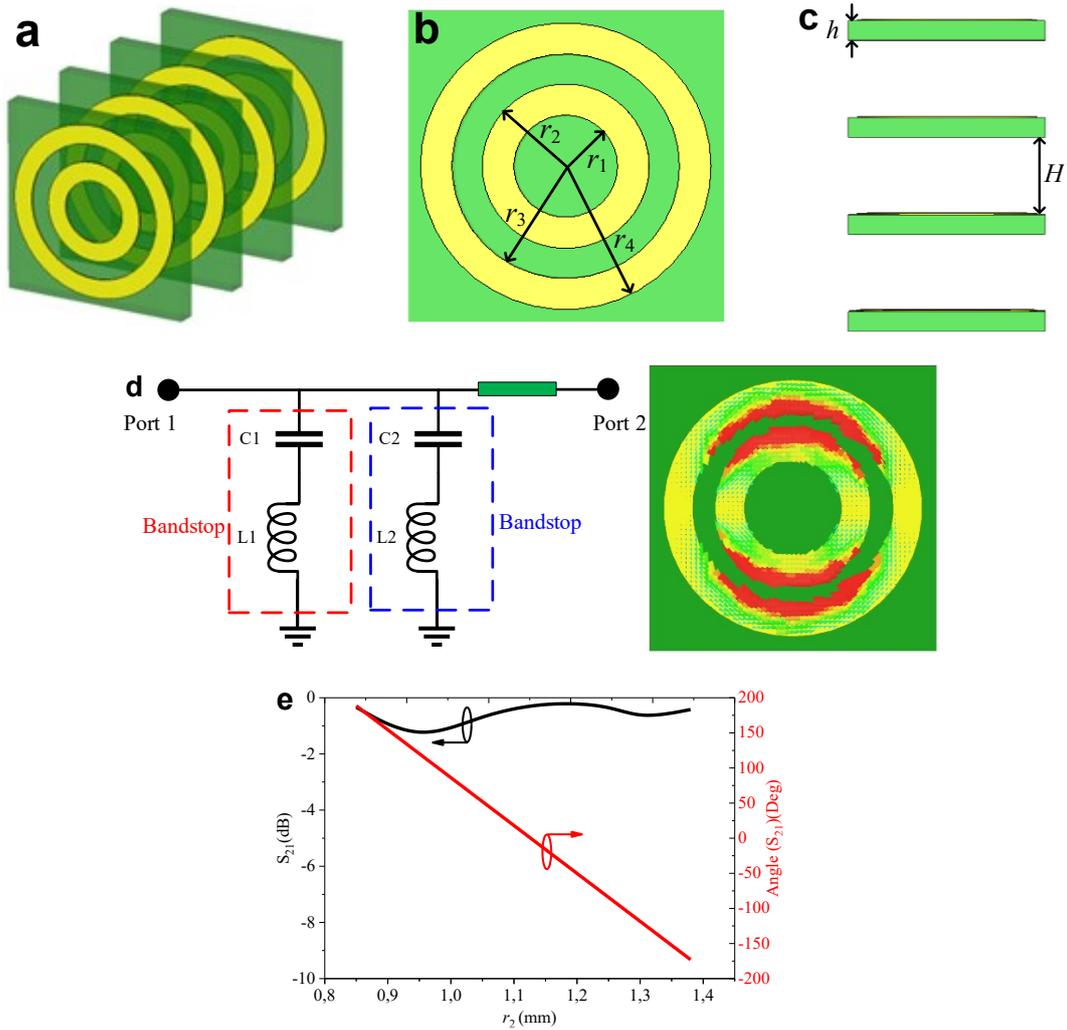

**Supplementary Fig. S6 | The geometry of the meta-atom and its frequency response. a** Perspective view of the meta-atom. **b** Front view of the meta-atom. **c.** Side view of the meta-atom. **d** The equivalent circuit of the single layer and current distribution on the double metallic rings at the frequency reflection zero. **e** Transmission amplitude and transmission phase of the meta-atom with different values of $r_2$ at 28 GHz. ($r_4$ = 2.3 mm, $r_3$ = 1.8 mm, $r_2 - r_1$ = 0.5 mm, H = 2.0 mm, h = 0.508 mm)